\documentclass[reprint,  longbibliography, footinbib, amsmath, amssymb, aps, pra]{revtex4-2}
\pdfoutput=1
\usepackage[utf8]{inputenc}
\usepackage{graphicx}
\usepackage{dcolumn}
\usepackage{bm}
\usepackage[dvipsnames]{xcolor}
\usepackage[colorlinks=true,linkcolor=teal,citecolor=Maroon,urlcolor=Maroon]{hyperref}
\usepackage{physics}
\usepackage{siunitx}
\usepackage{float}
\usepackage{tikz}
\usetikzlibrary{quantikz}
\usepackage{mathbbol}
\usepackage{enumitem}

\newcommand{\qgate}[1]{\textsc{#1}}

\newcommand{\SCQ}[2]{\mathcal{C}^{#1}_{#2}}
\newcommand{\GBF}{\Gamma_{\text{bit-flip}}}
\newcommand{\GPF}{\Gamma_{\text{phase-flip}}}

\newcommand{\secref}[1]{Section~\hyperref[#1]{\ref*{#1}}}
\newcommand{\appref}[1]{Appendix~\hyperref[#1]{\ref*{#1}}}
\newcommand{\tabref}[1]{Table~\hyperref[#1]{\ref*{#1}}}
\newcommand{\figref}[1]{Fig.~\hyperref[#1]{\ref*{#1}}}
\newcommand{\sfigref}[2]{Fig.~\hyperref[#1]{\ref*{#1}(#2)}}

\usepackage{svg}
\newcommand{\orcid}[1]{\href{https://orcid.org/#1}{\includegraphics[width=8pt]{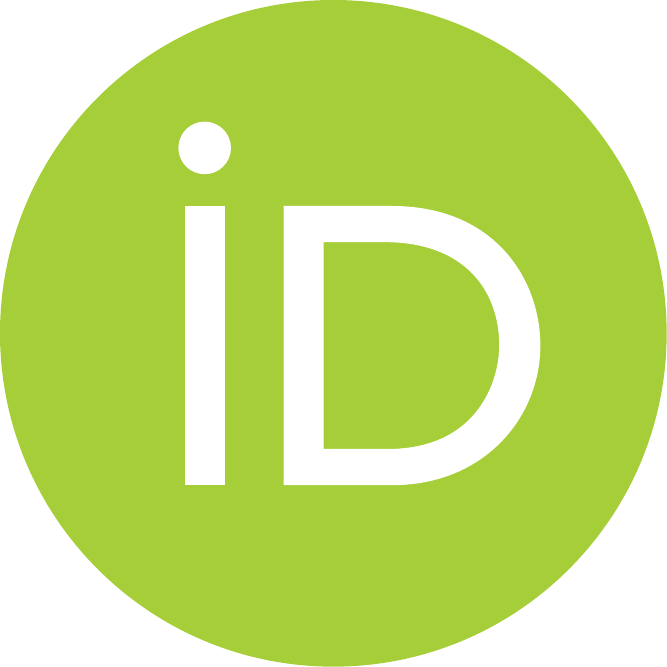}}}
\newcommand{\timo}[1]{{#1}}
\newcommand{\fernando}[1]{{#1}}

\begin{document}

\title{Quantum error correction with dissipatively stabilized squeezed cat qubits}
\author{Timo Hillmann\orcid{0000-0002-1476-0647}}
\email{timo.hillmann@rwth-aachen.de}
\affiliation{Department of Microtechnology and Nanoscience (MC2), Chalmers University of Technology, SE-412 96 Gothenburg, Sweden}
\author{Fernando Quijandría\orcid{0000-0002-2355-0449}}
\affiliation{Quantum Machines Unit, Okinawa Institute of Science and Technology Graduate University, Onna-son, Okinawa 904-0495, Japan}

\date{\today}

\begin{abstract}
Noise-biased qubits are a promising route toward significantly reducing the hardware overhead associated with quantum error correction. 
The squeezed cat code, a non-local encoding in phase space based on squeezed coherent states, is an example of a noise-biased (bosonic) qubit with exponential error bias.
Here we propose and analyze the error correction performance of a dissipatively stabilized squeezed cat qubit.
We find that for moderate squeezing the bit-flip error rate gets significantly reduced in comparison with the ordinary cat qubit while leaving the phase flip rate unchanged.
Additionally, we find that the squeezing enables faster and higher-fidelity gates.
\end{abstract}
\maketitle

\section{Introduction}
The interaction of a quantum system with its environment leads to the loss of quantum coherence. By tailoring the coupling of a quantum system to its environment, typically through an ancilla, well-established reservoir engineering methods
allow overcoming the decoherence paradigm 
by creating an effective dissipative dynamics which evolves in the long time to a target quantum state or a manifold of quantum states~\cite{poyatos_quantum_1996, krauter_entanglement_2011,  murch_cavity-assisted_2012, shankar_autonomously_2013, kienzler_quantum_2015, dassonneville_dissipative_2021}. 

In particular, in the field of circuit quantum electrodynamics (cQED)~\cite{blais_circuit_2021}, reservoir engineering has been successfully exploited to autonomously protect quantum information encoded in the infinite Hilbert space of a harmonic oscillator, i.e., a bosonic code, 
without the need of measurement-based feedback. 
This is achieved through the engineering of an effective parity-preserving dissipative evolution which drives the state of a microwave resonator to a manifold spanned by even and odd parity coherent superpositions of coherent states with opposite displacements also known as Schrödinger cat states~\cite{mirrahimi_dynamically_2014,leghtas_confining_2015, touzard_coherent_2018, lescanne_exponential_2020}. 
In principle, these dissipative dynamics could be used to prepare the logical states of the cat code~\cite{leghtas_confining_2015}.
Nevertheless, this is not necessary as universal control of a microwave resonator field using a dispersively coupled qubit is possible using optimal control pulse sequences~\cite{touzard_coherent_2018} or as it has been recently demonstrated, optimized sequences of continuous variables (CV) universal gate sets~\cite{eickbusch_fast_2022,kudra_robust_2022}. Therefore, reservoir engineering is left for the sole purpose of 
stabilizing 
the cat code.

Superpositions of squeezed vacuum states were introduced by Sanders~\cite{sanders_superposition_1989}.
Later, 
Hach III and Gerry~\cite{hach_iii_properties_1993} and Xin et al.~\cite{xin_even_1994} 
studied the nonclassical properties of coherent superpositions of squeezed states.
The latter are the states that result from the sequential application of displacement and squeezing operations on the photon vacuum with the squeezed vacuum state corresponding to the special case of zero displacement.
In particular, in this work, we will focus on the so-called \textit{squeezed cat states}. These are generalizations of the ordinary cat states and correspond to coherent superpositions of squeezed states with displacements of opposite amplitude and equal squeezing. The main interest in these states was spawned by the fact that
they actually represent superpositions of macroscopic quantum states as opposed to cat states which correspond to superpositions of nearly classical states.
Squeezed cat states were first realized in the optical domain through breeding and heralding detection operations~\cite{etesse_experimental_2015, huang_optical_2015}. 
In Ref.~\cite{lo_spinmotion_2015} entangled states of two displaced squeezed states of motion and the spin degrees of freedom of a trapped ion were realized. This work already highlighted the potential of these states for metrology.
\fernando{Later, Knott et al.~\cite{knott_practical_2016} demonstrated that squeezed cat states provided an advantage for sensing in the low-photon regime as compared to more conventional CV states. }

Recently, Schlegel et al. introduced the squeezed cat bosonic code~\cite{schlegel_quantum_2022}. 
This is the squeezed counterpart of the ordinary cat code in which logical states correspond to squeezed cat states.
Contrary to the cat code, in the squeezed cat code it is possible to approximately satisfy the Knill-Laflamme conditions for both single-photon loss and dephasing errors simultaneously in the large squeezing limit as well as the large coherent displacement limit.
In other words, the squeezed cat code merges the most notable quantum error correction features of both, cat and Gottesman-Kitaev-Preskill (GKP) codes, namely, 
the ability to correct pure dephasing and single-photon loss errors, respectively~\cite{gottesman_encoding_2001,albert_performance_2018, joshi_quantum_2021}.

In this work, we study the error correction potential of a squeezed cat qubit under a dissipative stabilization scheme which 
confines the state of the harmonic oscillator to the squeezed cat qubit manifold. This mechanism is a generalization of the cat qubit confinement~\cite{hach_iii_generation_1994, mirrahimi_dynamically_2014} and here we provide a possible implementation using superconducting circuits.
While the results presented in~\cite{schlegel_quantum_2022} indicate an increased performance of the squeezed cat code for an optimal recovery operation, their chosen metric, the average channel fidelity~\cite{nielsen_simple_2002}, does not distinguish between bit- and phase-flip errors. Because the squeezed cat qubit represents a biased noise qubit, we independently evaluate bit- and phase-flip errors
in the presence of single photon losses, photon gain, and pure photon dephasing. In addition to these decoherence processes, we also consider the effect of a residual Kerr interaction.

The subsequent sections of the article are organized as follows.
We begin in \secref{sec:definitions} by reviewing relevant properties of the squeezing and displacement operations and introduce the necessary notation.
Then, in \secref{sec:dissipative_stabilization}, we describe a theoretical framework that allows the dissipative stabilization of coherent superpositions of Gaussian states from which the stabilization scheme for squeezed cat states is derived.
In \secref{sec:squeezed_cat_qubit} we utilize the aforementioned stabilization scheme to analyze the error correction capabilities of the squeezed cat qubit. 
To this end, we introduce the squeezed cat code in \secref{ssec:squeezed_cat_code} before presenting the main results of this article in \secref{sec:numerical_results}.
Our findings show an exponential (in terms of the peak squeezing) reduction of the bit-flip error rate in comparison with the ordinary cat qubit without affecting phase-flip error rates.
However, at the same time, our results also highlight the need for small residual (Kerr) nonlinearities, as is the case for the GKP code as well.
The performance of the single qubit $Z$ gate is evaluated as well, suggesting exponentially faster and less noisy gates for the squeezed cat qubit.
To pave the way towards an experimental implementation, we propose in \secref{ssec:circuit_qed_implementation} a superconducting circuit based on Ref.~\cite{lescanne_exponential_2020} that realizes the dissipative stabilization scheme.
We conclude the article with a discussion of the results in \secref{sec:discussion}.
 
\section{Definitions \label{sec:definitions}}
We are going to restrict to a single mode bosonic field with annihilation (creation) operator $\hat a$ ($\hat a^\dagger$) obeying the commutation relation $[\hat a, \hat a^\dagger] = 1$.
The unitary displacement operator is defined by
\begin{equation}\label{eq:displace}
    \hat D(\alpha) = \exp \left( \alpha \hat a^\dagger - \alpha^* \hat a \right) ,
\end{equation}
and the unitary squeezing operator is defined by
\begin{equation}\label{eq:squeeze}
    \hat S(\xi) = \exp \left[ \frac{1}{2}\left( \xi^* \hat a^2 - \xi \hat a^{\dagger 2} \right) \right] ,
\end{equation}
with $\xi = r {\rm e}^{i \phi}$.
Their action on the annihilation operator $\hat a$ is given by
\begin{equation}\label{eq:displace-transformation}
    \hat D^{\dagger} (\alpha) \, \hat a \, \hat D(\alpha) = \hat a + \alpha ,
\end{equation}
and
\begin{equation}\label{eq:squeeze-transformation}
    \hat S^{\dagger} (\xi) \, \hat a \, \hat S(\xi) = \cosh(r) \,\hat a - {\rm e^{-i \phi}} \sinh(r) \, \hat a^\dagger
\end{equation}
respectively.

Following Refs.~\cite{lu_new_1971, hollenhorst_quantum_1979, walls_squeezed_1983}
a \emph{squeezed state} $\ket{\alpha, \xi}$ (also squeezed-coherent or squeezed-displaced state)
is the state that results from the sequential application of the squeezing operator~\eqref{eq:squeeze} and the displacement operator~\eqref{eq:displace} on the photon vacuum state
\begin{equation}\label{eq:squeezed-state}
    \ket{\alpha, \xi} = \hat D(\alpha) \hat S(\xi)\ket{0} .
\end{equation}
The $\alpha = 0$ case corresponds to the well-known squeezed vacuum state. 
An alternative definition of a squeezed state was given by Yuen~\cite{yuen_two-photon_1976}. 
This state is called the \emph{two-photon coherent state} $\ket{\alpha}_\xi$ and it is defined by first displacing the vacuum state and then squeezing it
\begin{equation}\label{eq:two-photon-coherent}
    \ket{\alpha}_\xi \equiv \hat S(\xi) \, \hat D (\alpha) \ket{0} .
\end{equation}
From the relations \eqref{eq:displace-transformation} and \eqref{eq:squeeze-transformation} it is straightforward to show that
\begin{align}
    \hat D(\alpha) \hat S(\xi) &= \hat S(\xi) \hat D \left[ \alpha \cosh(r) + \alpha^* {\rm e}^{-i\phi} \sinh(r) \right] , 
\end{align}
which establishes the relation between squeezed and two-photon coherent states
\begin{equation}\label{eq:equivalence}
    \ket{\alpha, \xi} \equiv \ket{\alpha \cosh(r) + \alpha^* {\rm e}^{-i\phi} \sinh(r)}_\xi .
\end{equation}
In this work, we are going to stick to the squeezed states as defined by Eq.~\eqref{eq:squeezed-state}.

The squeezed cat states are defined as the coherent superposition of two squeezed cat states with opposite displacement amplitudes and identical squeezing
\begin{equation}\label{eq:squeezed_cat_code}
    \ket*{\mathcal{C}^{\pm}_{\alpha, \xi}} = \frac{1}{N^\pm_{\alpha, \xi}} \left( \ket{\alpha, \xi} \pm \ket{-\alpha, \xi} \right), 
\end{equation}
where $N_{\alpha, \xi}^{\pm}$ is a normalization constant. These states can be thought of as generalizations of the cat states. Similarly, these are parity eigenstates with $\ket*{\mathcal{C}^{+}_{\alpha, \xi}}$ ($\ket*{\mathcal{C}^{-}_{\alpha, \xi}}$) a superposition of even (odd) number states. This property makes them suitable candidates for designing a bosonic code as studied in Ref.~\cite{schlegel_quantum_2022}. 

\section{Dissipative stabilization of coherent superpositions of Gaussian states \label{sec:dissipative_stabilization}}
Here we build on the result by Hach III and Gerry~\cite{hach_iii_generation_1994}. 
Consider a single-mode bosonic system whose non-unitary dynamics are described by a Gorini-Kossakowski-Sudarshan-Lindblad master equation~\cite{gorini_completely_1976, lindblad_generators_1976} of the form (we set $\hbar = 1$ throughout this paper)
\begin{equation}\label{eq:Lindblad-general}
    \dv{\hat{\rho}}{t} = -i [\Omega \hat L^\dagger + \Omega^* \hat L, \rho] + 
    \kappa \mathcal{D}[\hat{L}] \hat{\rho},
\end{equation}
where $\mathcal{D}[\hat{A}] \hat{\rho} = \hat{A} \hat{\rho} \hat{A}^{\dagger} - \frac{1}{2} \hat{A}^{\dagger} \hat{A} \hat{\rho} - \frac{1}{2} \hat{\rho} \hat{A}^{\dagger} \hat{A}$ and 
where the operator $\hat L$ is, in general, a function of the bosonic annihilation ($\hat a$) and creation ($\hat a^\dagger$) operators. Then, the steady state $\partial_t \hat{\rho}_{\rm ss} = 0$ of \eqref{eq:Lindblad-general} is an eigenstate of the operator $\hat L$ with eigenvalue $z = - 2 i \Omega / \kappa$, i.e.,
\begin{equation}\label{eq:ss-condition}
    \hat L \hat{\rho}_{\rm ss} = z \hat{\rho}_{\rm ss}.
\end{equation}
This allows one to express Eq.~\eqref{eq:Lindblad-general} in a very concise form 
\begin{align}
    \dv{\hat{\rho}}{t} = \kappa \mathcal{D}(\hat{L} - z)\hat{\rho} .
\end{align}

Following Eq.~\eqref{eq:ss-condition}, the most general steady-state of \eqref{eq:Lindblad-general} would be a statistical mixture of eigenstates of $\hat L$ with a common eigenvalue.
This is the basis of the dissipative stabilization of cat states~\cite{mirrahimi_dynamically_2014, leghtas_confining_2015, touzard_coherent_2018}. 
A similar approach has been proposed for the stabilization of cat states in atomic ensembles~\cite{qin_generating_2021}.

Now, starting from $\hat D(\alpha) \hat S(\xi) \hat a \ket{0} = 0$, it is straightforward to show the relation
\begin{equation} \label{eq:squeezed-eigenvalue} 
    \hat b \ket{\alpha ; \xi} = \beta_{\alpha, \xi} \,
    \ket{\alpha ; \xi} ,
\end{equation}
where we have introduced the bosonic operator
$\hat b = \hat S (\xi) \, \hat a \, \hat S^{\dagger}(\xi) = \cosh(r) \,\hat a + {\rm e^{-i \phi}} \sinh(r) \, \hat a^\dagger$, and the related complex eigenvalue $\beta_{\alpha, \xi} = \alpha \cosh(r) +  \alpha^* {\rm e}^{-i \phi} \sinh(r)$.
From Eq.~\eqref{eq:squeezed-eigenvalue} the relation
$\hat b^n \ket{\alpha ; \xi} = \beta_{\alpha, \xi}^n \ket{\alpha ; \xi}$ for an arbitrary integer $n$ immediately follows.
In turn, from this relation, it follows that the squeezed cat states \eqref{eq:squeezed_cat_code} are degenerate eigenstates of the operator $\hat b^2$
\begin{equation}
    \hat b^2 \ket*{\mathcal{C}^{\pm}_{\alpha, \xi}} = \beta_{\alpha, \xi}^2 \ket*{\mathcal{C}^{\pm}_{\alpha, \xi}} .
\end{equation}
Similarly to the case of (Schrödinger) cat states, higher-order superpositions of squeezed states may yield higher-order powers of the eigenvalue $\beta_{\alpha, \xi}$. Nevertheless, in this work, we are going to restrict to the case $n=2$.

Following the above discussion, for $\hat L = \hat b^2$ the steady state of the dissipative dynamics will be, in general, a mixture of even and odd parity squeezed cat states.
However, as in this case photons are created and annihilated in pairs, the parity of an initial state will be preserved throughout the dynamics. In other words, an initial even (odd) state will evolve in the long time to the state $\ket*{\mathcal{C}^+_{\alpha, \xi}}$ ($\ket*{\mathcal{C}^{-}_{\alpha, \xi}}$).

For convenience, in this work we will restrict to the case of a superposition of two squeezed states with squeezing $\xi$ along the $x$ quadrature, i.e., $\phi =0$, and displaced along the $x$ axis, i.e., $\alpha$ real. In this case, $\beta_{\alpha, \xi}$ reduces to $\beta_{\alpha, \xi} = \alpha \exp(r)$.
By setting the drive amplitude $\Omega = i \bar{\Omega}$, with $\bar{\Omega}$ real, we can fix the steady-state eigenvalue $z =  2 \bar{\Omega} / \kappa$ to be real as well. Therefore, in order to stabilize a coherent superposition of $x$ squeezed states displaced along the $x$ axis, we choose the drive amplitude to be $\bar{\Omega} = \kappa \alpha^2 \exp(2 r)  /2$. 


\section{Application: Squeezed Cat Qubit \label{sec:squeezed_cat_qubit}}
\subsection{Introduction \label{ssec:squeezed_cat_code}}
\begin{figure*}
    \centering
    \includegraphics{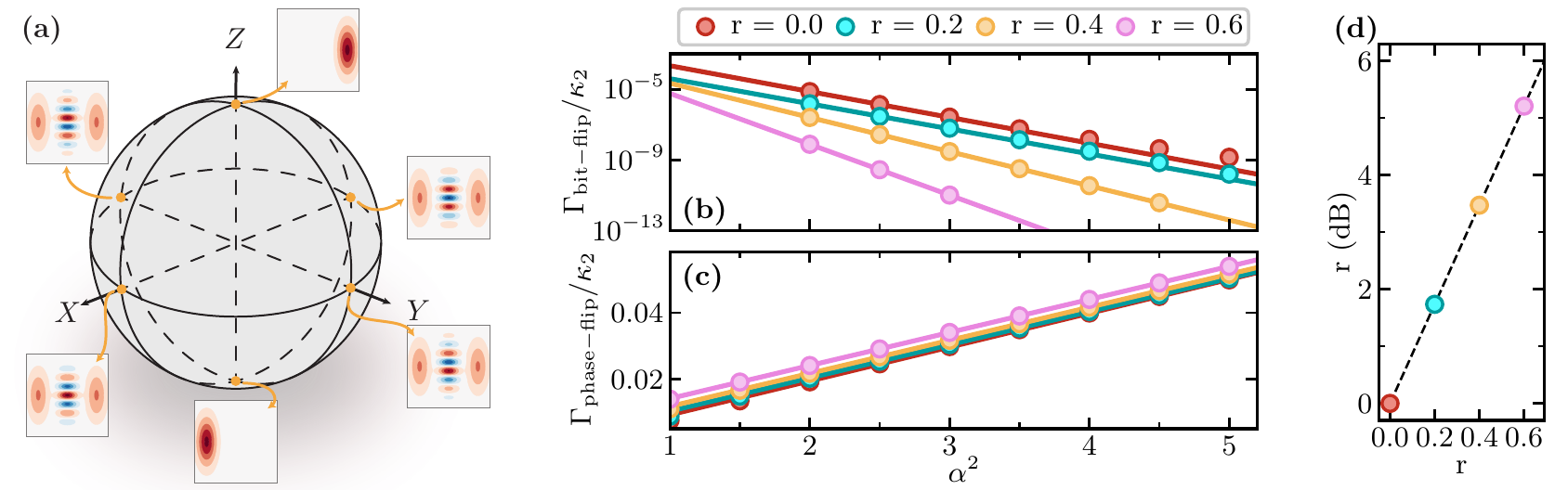}
    \caption{\textbf{(a)} Sketch of the squeezed cat code on the Bloch sphere.  \textbf{(b)} -  \textbf{(c)} Effective bit- and phase-flip rates for squeezed cat codes with different squeezing. Here we choose $\kappa_{-} / \kappa_2 = 10^{-3}$.
    Markers show the rates obtained from numerical simulations, while solid lines show exponential and linear fits in the range $2 \leq \lvert \alpha \rvert^2 \leq 5$ for panels \textbf{(b)} and \textbf{(c)}, respectively.
    \timo{We show results only for $\GBF \geq \num{e-13}$ due to numerical accuracy and stability issues for smaller rates.}
    \textbf{(d)} Visual representation of the conversion from the dimensionless
    squeezing parameter $r$ to experimentally relevant squeezing values in $\si{\dB}$.  }
    \label{fig:squeezed_cat_code}
\end{figure*}

We have seen above that the dynamics of an oscillator described by the Lindblad equation
\begin{align}
    \label{eq:ideal_cat_code_dissipator}
    \dv{\hat{\rho}}{t} = \kappa_{2} \mathcal{D}[\hat{b}^2 - \beta_{\alpha, r}^2],
\end{align}
with $\hat{b} = \hat{S}(r) \hat{a} \hat{S}^{\dagger}(r)$, $\hat{a}$ the annihilation operator of the oscillator mode and $\hat{S}(r)$ the squeezing operator, are restricted to the two-dimensional subspace spanned by the orthogonal squeezed cat states $\{ \ket*{\SCQ{+}{\alpha, r}}, \ket*{\SCQ{-}{\alpha, r}} \}$ with $\beta_{\alpha, r} = \alpha \mathrm{e}^{r}$.
This motivates the definition of the squeezed cat-qubit (SCQ) logical basis as
\begin{align}
    \ket*{\SCQ{0}{\alpha, r}} &= \frac{1}{\sqrt{2}} \left( \ket*{\SCQ{+}{\alpha, r}} +  \ket*{\SCQ{-}{\alpha, r}} \right) \approx \ket{\alpha, r} 
    \label{eq:scq_zero} \\
     \ket*{\SCQ{1}{\alpha, r}} &= \frac{1}{\sqrt{2}} \left( \ket*{\SCQ{+}{\alpha, r}} -  \ket*{\SCQ{-}{\alpha, r}} \right) \approx \ket{-\alpha, r} 
    , \label{eq:scq_one}
\end{align}
where the approximation sign occurs because in contrast to $\ket*{\SCQ{\pm}{\alpha, r}}$, the squeezed coherent states $\ket{\pm\alpha, r}$ are only quasi-orthogonal, that is, their finite overlap is given by
\begin{align}
    \braket{-\alpha, r}{\alpha, r} = \exp(- 2 \alpha^2 \mathrm{e}^{2 r}).
\end{align}
A Bloch sphere representation of the SCQ is shown in \figref{fig:squeezed_cat_code}(a).
The squeezed cat code is related to the ordinary cat code in the sense that one recovers the ordinary cat code from the squeezed version in the limit of zero squeezing $(r \rightarrow 0)$.
In the opposite limit, $r \rightarrow +\infty$, the code becomes translation invariant with respect to phase space translations of amplitude $s = 2\pi k / \abs{\alpha}, k \in \mathbb{Z}$ along the $p$ quadrature.
The translation invariance in this limit relates the code to the GKP code. In fact, one can interpret the squeezed cat code as a (bad) approximation of the GKP code that has only two (infinitely) squeezed peaks.
This view explains intuitively the finite error correction capabilities of the SCQ against phase flips the authors of Ref.~\cite{schlegel_quantum_2022} found through an analysis of the Knill-Laflamme conditions.
From the perspective of the ordinary cat code, 
the increased error correction capabilities arise as the squeezed coherent state is not an eigenstate of the annihilation operator anymore such that \timo{for $r > 0$ the state $\hat{a}\ket*{\SCQ{\pm}{\alpha, r}}$ has a finite component that is orthogonal to the code space, i.e., it lies outside of the code space in the error space.
}

In Ref.~\cite{schlegel_quantum_2022} the authors also demonstrate numerically the increased error correction performance of the squeezed cat qubit over the ordinary cat qubit by computing and applying the optimal recovery operation obtained from a semidefinite program.
While one can argue that the optimal recovery operation allows one to compute the maximally achievable performance of a given quantum code, physically implementing the required recovery is in many cases non-trivial.
Thus, dissipative stabilization schemes such as the one described here are typically more practical and belong to the class of confinement schemes that counteract the leakage of states out of the code space.
For the ordinary cat code, dissipative~\cite{guillaud_repetition_2019, mirrahimi_dynamically_2014, touzard_coherent_2018, lescanne_exponential_2020}, Hamiltonian~\cite{puri_stabilized_2019, puri_engineering_2017, grimm_stabilization_2020} and combined~\cite{gautier_combined_2022} confinement schemes have been analyzed.
In \appref{app:kerr_squeezed_cat} we give reasons why Hamiltonian confinement schemes realized in superconducting circuits are impractical over the dissipative scheme considered here.

\subsection{Main Results \label{sec:numerical_results}}
Here we analyze numerically the error correction capabilities of 
the dissipative confinement into the squeezed cat code manifold of an initially prepared ideal SCQ state
by assessing the suppression of bit-flip errors and scaling of phase-flip errors.
The master equation describing the confinement dynamics together with typical decoherence effects is
\begin{align}
    \label{eq:master_eq_numerical_results}
    \begin{split}
    \dv{\hat{\rho}}{t} =& -i \comm*{\hat{H}_{\mathrm{Kerr}}}{\hat{\rho}}  + \kappa_2 \mathcal{D}(\hat{b}^2 - \beta_{\alpha, r}^2) \hat{\rho} \\ 
    +& \kappa_{-} \mathcal{D}[\hat{a}]\hat{\rho} + \kappa_{\phi} \mathcal{D}[\hat{a}^{\dagger}\hat{a}]\hat{\rho} + \kappa_{+} \mathcal{D}[\hat{a}^{\dagger}]\hat{\rho} ,
    \end{split}
\end{align}
where $\kappa_{-} = \kappa_1 (1 + n_{\rm th}) $, $\kappa_{\phi}$, and $\kappa_{+} = \kappa_{1} n_{\rm th}$ denote the rates of photon loss, pure  photon number dephasing, and photon gain, respectively with $n_{\rm th}$ the mean number of thermal excitations and $\kappa_1$ the spontaneous emission rate.
The rate $\kappa_2$ determines the dissipative confinement rate and $\hat{H}_{\mathrm{Kerr}} = K \hat{a}^{\dagger 2} \hat{a}^2$ is the Kerr Hamiltonian.
We choose the effective bit- and phase-flip rates $\GBF$ and $\GPF$ as a metric to evaluate the performance of the SCQ which are obtained from the time-evolution of the state $ \ket*{\SCQ{0}{\alpha, r}}$ and $ \ket*{\SCQ{+}{\alpha, r}}$, respectively.
These rates describe the time scale on which the expectation values of the logical SCQ operators $\expval{\hat{\sigma}_Z(t)}$ and $\expval{\hat{\sigma}_X(t)}$ decay in the presence of different decoherence processes, that is, $\expval{\hat{\sigma}_Z(t)} \sim \mathrm{e}^{- \GBF t}$ and $\expval{\hat{\sigma}_X(t)} \sim  \mathrm{e}^{- \GPF t}$.
Additional details about the explicit construction of the observables $\hat{\sigma}_{Z}$ and $\hat{\sigma}_{X}$, as well as our numerical methods, can be found in \appref{app:numerical_methods}.
Furthermore, to avoid overloading notation, from here on we make the dependence of $\beta_{\alpha, r}$ on $\alpha$ and $r$ implicit and instead write $\beta$.

\subsubsection{Bit- and phase-flip error rates with single-photon losses}
As already stated in the Introduction, the SCQ represents a biased-noise qubit and therefore, we will independently evaluate bit- and phase-flip errors.

Since we consider the implementation of the dissipative stabilization scheme in superconducting circuits, we begin by investigating the effects of single-photon losses while ignoring all other noise sources, that is, $\kappa_{-} \neq 0$ and $K = \kappa_{\phi} = \kappa_{+} = 0$.
Our numerical results for $\GBF$ and $\GPF$ are shown in~\sfigref{fig:squeezed_cat_code}{b} and~\sfigref{fig:squeezed_cat_code}{c}, respectively.
There we chose $\kappa_{-} = \num{5e-3} \kappa_2$, but note that in the regime where $\kappa_{-} \ll \kappa_{2}$ both error rates scale linearly with $\kappa_{-}$. i.e., $\Gamma_{\mathrm{err}} \propto \kappa_{-}$.
We show the error rates for various values of the dimensionless squeezing parameter $r$ which lie in an experimentally feasible regime, see also \sfigref{fig:squeezed_cat_code}{d} which visualizes the conversion from the dimensionless quantity $r$ to the experimentally relevant dimensional quantify in dB given by  $r \, (\si{\dB}) = 20 r / \log(10)$.
We point out some relevant observations from the numerically determined bit-flip rate shown in \sfigref{fig:squeezed_cat_code}{b}.
Importantly, we find that the bit-flip rate decreases monotonically for any value of $\vert \alpha \vert^2$.
However, the exponential suppression factor $\gamma$, defined such that $\GBF \propto e^{- \gamma \lvert \alpha \rvert^2}$, only increases once $r > 0.2$.
This can also be seen from the exponential fits in the range $2 \leq \lvert \alpha \rvert^2 \leq 5$ shown as solid lines which are roughly parallel for $r = 0$ and $r = 0.2$, but become steeper for any value of $r > 0.2$.
The reason is that for the ordinary cat qubit the exponential scaling factor $\gamma_{\text{cat}}$ is not constant, but (weakly) dependent on $\alpha$ with $2 \leq \gamma_{\text{cat}} \leq 4$.
Numerical simulations have shown~\cite{guillaud_error_2021} that the upper end of this range, i.e., $\gamma_{\text{cat}} \approx 4$, is typically achieved for $\vert \alpha \vert^2 \approx 2$. 
For larger coherent displacements, $\gamma_{\text{cat}}$ steadily reduces until it reaches $\gamma_{\text{cat}} \approx 2$ once $\vert \alpha \vert^2 \gtrapprox 5$.
Associating a squeezed cat qubit with an ordinary cat qubit with displacement amplitude $\beta = \alpha e^{r}$, 
the observation of constant (or even decreased) effective scaling factor $\gamma = \gamma_{\text{cat}} e^{2 r}$ can be explained if $\gamma_{\text{cat}}$ for $\lvert \alpha \rvert^2 > 2$ initially decays faster than $\exp(2r)$.

At the same time we see that the slope of the phase-flip rate $\GPF$ shown in \sfigref{fig:squeezed_cat_code}{c} does not increase with $r$ and $\GPF$ is approximately independent of $r$.
This behavior is predicted from the corresponding transition matrix element, i.e.,
\begin{align}
     \lvert \bra*{\SCQ{+}{\alpha, r}} \hat{a} & \ket*{\SCQ{-}{\alpha, r}} \rvert^2 \nonumber \\ 
     &= \beta^2 \left\lvert \cosh (r) \tanh(\beta^2) - \sinh (r) \coth(\beta^2) \right\rvert^2 \nonumber \\
     &\stackrel{\alpha e^r \gg 1}{\longrightarrow}\alpha^2,
\end{align}
Thus, it is possible to further increase the noise bias of the dissipatively stabilized cat qubit if we instead stabilize a SCQ with $r > 0$.
Alternatively, it is possible to keep the bit-flip rate $\GBF$ fixed while reducing the phase-flip rate by decreasing $\lvert \alpha \rvert^2$ and increasing $r$ suitably.

While we have seen now that the stabilized SCQ outperforms the stabilized cat qubit in the presence of single-photon losses, it is essential to establish that these advantages persist in a more general error model.
To this end, we analyze the bit-flip rate $\GBF$~\footnote{
We have verified numerically that the phase-flip rate remains independent of the squeezing parameter $r$ for the investigated decoherence processes.}
of the stabilized SCQ in the presence of pure dephasing, single-photon gain, and undesired coherent Kerr evolution in the following sections.

\subsubsection{Influence of pure dephasing noise}

\begin{figure}
    \centering
    \includegraphics{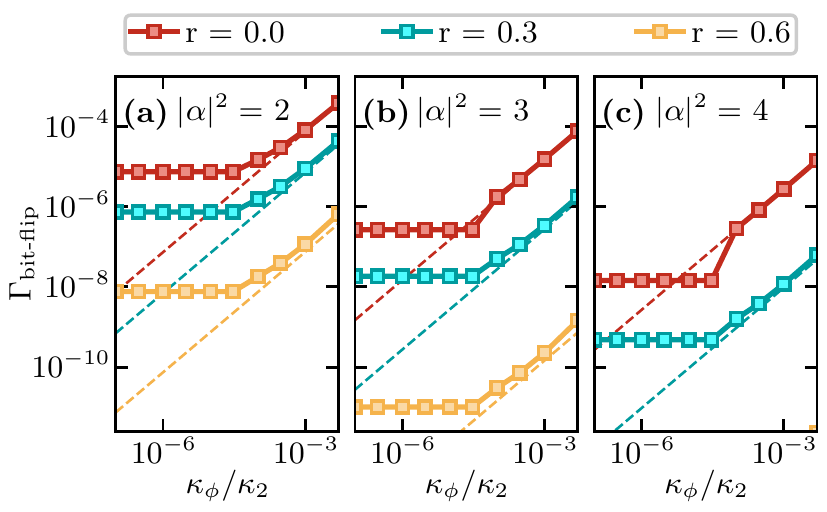}
    \includegraphics{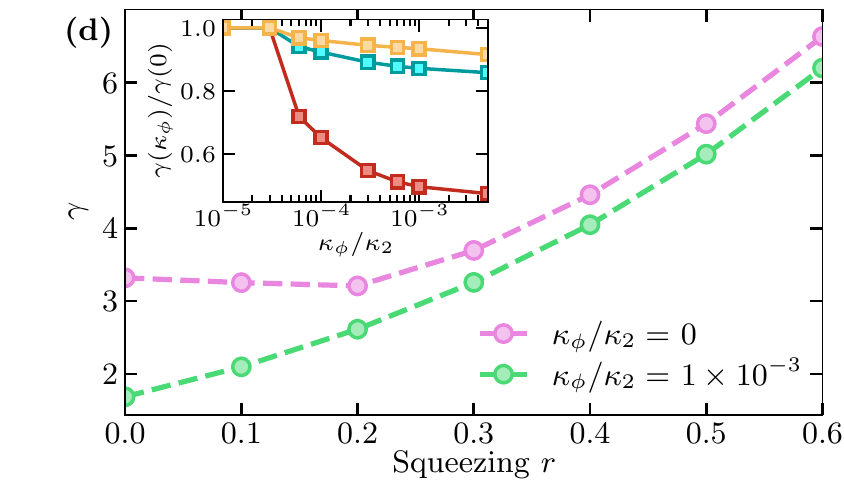}
    \caption{\textbf{(a) - (c)} Bit-flip rate as function of the pure dephasing rate $\kappa_{\phi}$ for different combinations of displacement amplitude $\alpha$ and squeezing parameter $r$. \timo{Note that we show results only for $\GBF \geq \num{e-13}$ due to numerical accuracy and stability issues for smaller rates.}\textbf{(d)} Exponential suppression factor of bit-flip errors $\gamma$ with $\GBF \propto \exponential(-\gamma \lvert \alpha \rvert^2)$ for the idling, dissipatively stabilized SCQ as a function of the squeezing $r$.
    The inset shows the fraction $\gamma(\kappa_{\phi}) / \gamma(0)$ as a function of $\kappa_{\phi}$ for selected values of $r$, indicating that for $r > 0$ the suppression factor becomes increasingly independent of $\kappa_{\phi}$ for the selected parameter range. The stabilized state is subject to single-photon losses with rate $\kappa_{-} = 5 \times 10^{-3} \kappa_2$ as well as pure dephasing with rates $\kappa_{\phi} = 0$ (purple) and $\kappa_{\phi}= \num{e-3} \kappa_2$ (green). In (d) the value of $\gamma$ is determined through an exponential fit in the range $2 \leq \lvert \alpha \rvert^2 \leq 5$.
    Connecting lines are a guide for the eye.}
    \label{fig:dephasing_suppression}
\end{figure}

Pure dephasing is a noise process that is commonly overlooked in theoretical studies of superconducting microwave cavities.
While it is usually a good assumption to neglect pure dephasing in a linear resonator, this assumption is not necessarily true once the resonator is (weakly) coupled to a nonlinear element, e.g., an auxiliary qubit. 
The interplay of spontaneous excitation of the qubit and the dispersive coupling between the qubit and resonator can then be described by an effective pure dephasing noise model if the qubit is traced out.
Additionally, number dephasing is also the result of the coupling of the storage mode to a flux tunable device, in general, a superconducting loop or loops interrupted by Josephson junctions. In this case, the sensitivity of this device to the magnetic flux noise is inherited by the storage mode with its frequency drifting stochastically in time which results in number dephasing. Nevertheless, this is a second-order effect as compared to the dephasing arising from the dispersive coupling to a qubit.
\timo{This motivates us to determine the parameter regime in which the effects of pure dephasing noise become relevant for the performance of the dissipatively stabilized (squeezed) cat qubit.}

\timo{For the ordinary cat qubit it has been shown that, if additionally to single-photon losses there is non-negligible pure dephasing, the exponential suppression factor $\gamma$ is expected to be constant and equal to 2~\cite{mirrahimi_cat-qubits_2016}.}
\timo{We find that for the SCQ pure dephasing effects become dominant once $\kappa_{\phi} / \kappa_{-} \geq \num{e-2}$ and that these are also well described by a model for the bit-flip rate with $\gamma = 2 e^{r}$ as we describe in more detail in the forthcoming paragraphs.}

Our results are summarized in \figref{fig:dephasing_suppression}. Here we show the effective rate $\GBF$ as a function of $\kappa_{\phi}$ for different values of $\alpha$ in \sfigref{fig:dephasing_suppression}{a-c} and the scaling of $\gamma$ as a function of the squeezing parameter $r$ in \sfigref{fig:dephasing_suppression}{d}.
The numerical data is obtained by simulating the evolution of the initial SCQ $\ket*{\SCQ{0}{\alpha, r}}$ undergoing the dissipative dynamics given by Eq.~\eqref{eq:Lindblad-general}
with single-photon loss rate $\kappa_{-} = 5 \times 10^{-3} \kappa_2$ for various pure dephasing rates $\kappa_{\phi}$ while $\kappa_{+} = K = 0$.
From this data the value of $\gamma$ shown in \sfigref{fig:dephasing_suppression}{d} is determined by an exponential fit of $\GBF$ in the range $2 \leq \lvert \alpha \rvert^2 \leq 5$.
There are a few noteworthy observations that can be made from \figref{fig:dephasing_suppression}.
First, we notice that $\GBF$ 
is only affected by pure dephasing 
for $\kappa_\phi / \kappa_- \gtrsim 10^{-2}$
for the chosen simulation parameters. 
However, we observe the transition point is independent of the code parameters $\alpha$ and $r$ and that the contribution of pure dephasing to the bit-flip rate ($\GBF^{(\kappa_{\phi})}$) is well approximated by
\begin{align}
    \label{eq:bit_flip_rate_dephasing_thy}
    \GBF^{(\kappa_{\phi})} &\approx \kappa_{\phi} \cosh^2(2r) \lvert \beta \rvert^2 \sinh^{-1}(2 \lvert \beta \rvert^2).
\end{align}
This approximation for $\GBF^{(\kappa_{\phi})}$ is obtained by the replacement $\alpha \to \beta$ in Eq.~(A.9) of Ref.~\cite{mirrahimi_dynamically_2014} and multiplying the expression by $\cosh^2(2r)$ which is a result of the squeezing transformation~\footnote{
In Ref.~\cite{mirrahimi_dynamically_2014} $\GBF^{(\kappa_{\phi})}$ is calculated from the matrix element $\bra*{C_\alpha^{-}}\mathcal{D}[\hat{a}^{\dagger} \hat{a}] J_{+-}^{\dagger}\ket*{ C_\alpha^{+}}$ where $ J_{+-}^{\dagger}$ is defined in Eq.~\eqref{eq:app_j_pm}.
For the SCQ  $\GBF^{(\kappa_{\phi})}$ can be calculated from $\bra*{C_\beta^{-}}\mathcal{D}[\hat{b}^{\dagger} \hat{b}] J_{+-}^{\dagger}\ket*{ C_\beta^{+}}$ where $\hat{b} = \cosh(r) \hat{a} + \sinh(r) \hat{a}^{\dagger}$.
Under the crude approximation $\mathcal{D}[\hat{b}^{\dagger} \hat{b}] \approx \mathcal{D}[\cosh(2r) \hat{a}^{\dagger} \hat{a}]$ one obtains Eq.~\eqref{eq:bit_flip_rate_dephasing_thy}. 
}.
A derivation of the exact expression of $\GBF^{(\kappa_{\phi})}$ for the SCQ is left as an open problem.
The predicted bit-flip rate $\GBF$ [Eq.~\eqref{eq:bit_flip_rate_dephasing_thy}] due to pure dephasing noise is shown as dashed lines in \sfigref{fig:dephasing_suppression}{a-c}.
The existence of the above mentioned transition point also becomes apparent in the inset of \sfigref{fig:dephasing_suppression}{d} which shows the exponential suppression factor $\gamma(\kappa_{\phi})$ normalized to its value at $\kappa_{\phi} = 0$ for different values of $r$.
The inset and main panel of \sfigref{fig:dephasing_suppression}{d} also show that $\gamma$ is only weakly dependent of $\kappa_{\phi}$ for $r > 0.2$.
In the main panel, we have also shown the behavior of $\gamma$ as a function of $r$ in the absence of pure dephasing noise (purple markers).
For $r \leq 0.2$, $\gamma$ stays approximately constant after which it increases exponentially with rate $\approx e^{2 r}$, in agreement with the observation in \sfigref{fig:squeezed_cat_code}{b}.

\timo{
\subsubsection{Influence of single-photon gain}
As noted in the beginning of \secref{sec:numerical_results}, 
the single-photon excitation rate of the system
is given by $\kappa_{+} = n_{\text{th}} \kappa_{1}$.
Typically, the effective temperature of the bath to which the system couples is such that $n_{\text{th}} \approx \num{e-2}$.
The action of the creation operator $\hat{a}^\dagger$ on a cat qubit state will lead to leakage outside of the cat qubit code space into an excited state manifold. 
Even though the engineered two-photon dissipation mechanism will correct for this leakage, while in the excited state manifold, the probability of tunneling into the other potential well is increased.
Considering that the height of the potential well scales with $\alpha$ for the ordinary cat qubit~\cite{lescanne_exponential_2020}, we expect that the SCQ should be less affected by thermal noise due to its increased effective coherent displacement amplitude $\beta = \alpha e^{r}$.
%

\begin{figure}[b]
    \centering
    \includegraphics{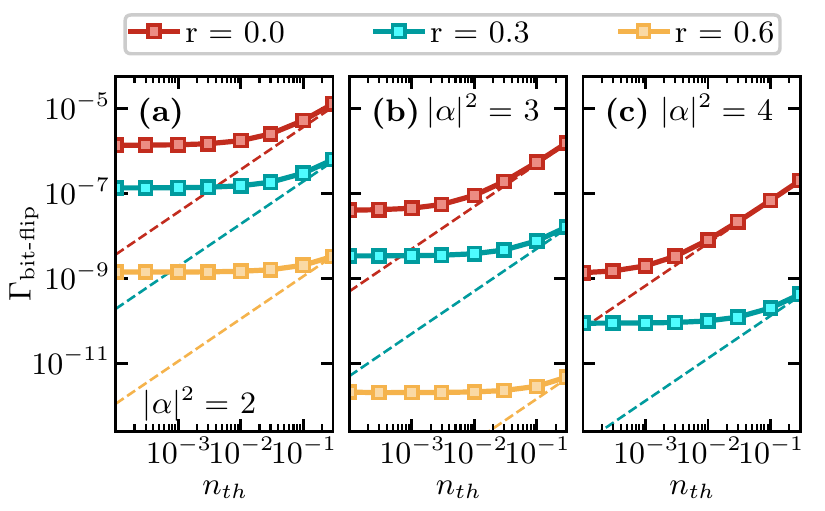}
    \includegraphics{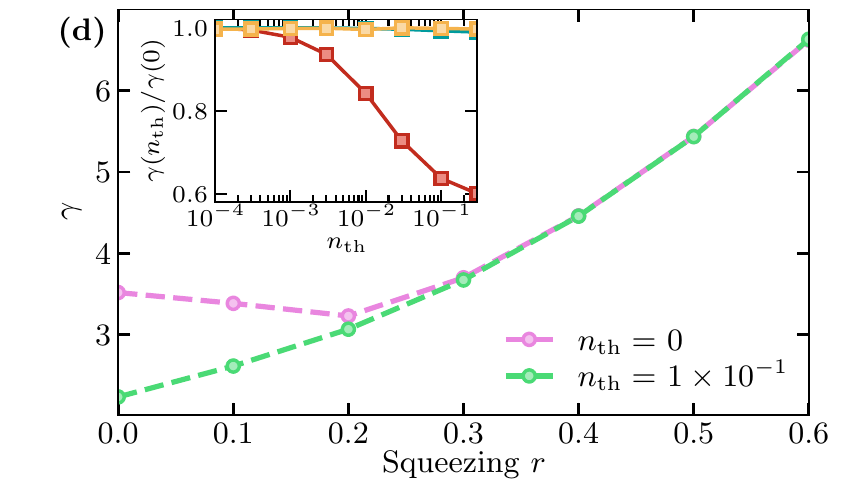}
    \caption{\timo{ \textbf{(a) - (c)} Bit-flip rate as a function of the mean number of thermal excitations $n_{\text{th}}$ for different combinations of the displacement amplitude $\alpha$ and the squeezing parameter $r$. The spontaneous emission rate is chosen as $\kappa_1 / \kappa_2 = \num{e-3}$.
    Note that we show results only for $\GBF \geq \num{e-13}$ due to numerical accuracy and stability issues for smaller rates.
    \textbf{(d)} Exponential suppression factor of bit-flip errors $\gamma$ with $\GBF \propto \exponential(-\gamma \lvert \alpha \rvert^2)$ for the idle, dissipatively stabilized SCQ as a function of the squeezing $r$.
    See \figref{fig:dephasing_suppression} for a description of the parameter regime.
    The inset shows the fraction $\gamma(n_{\text{th}}) / \gamma(0)$ as a function of $n_{\text{th}}$ for selected values of $r$, indicating that for $r > 0$ the suppression factor becomes increasingly independent of $n_{\text{th}}$ for the selected parameter range. Connecting lines are a guide for the eye.}}
    \label{fig:gain_suppression}
\end{figure}

We now analyze the influence of single-photon gain in the bit-flip rate in a similar manner as we did for the case of pure dephasing in the previous section.
For our numerical simulations we have chosen $\kappa_{1}/\kappa_{2} = \num{e-3}$ while $\kappa_{\phi} = K = 0$.
Using the same approximation that lead to Eq.~\eqref{eq:bit_flip_rate_dephasing_thy} for $\GBF^{(\kappa_{\phi})}$, we obtain the following expression for the bit-flip rate due to single-photon gain, i.e.,
\begin{align}
    \label{eq:bit_flip_rate_gain_thy}
    \GBF^{(\kappa_{+})} &\approx \kappa_{+} \cosh^2(2r) \sinh^{-1}(2 \lvert \beta \rvert^2).
\end{align}
The contribution of $\GBF^{(\kappa_{+})}$ to the total bit-flip rate $\GBF$ is shown as dashed lines in \sfigref{fig:gain_suppression}{a-c}.
Comparing the different panels, it becomes apparent that there does not exist a single transition point in terms of the mean number of thermal excitations $n_{\text{th}}$ after which the contribution of $\GBF^{(\kappa_{+})}$ to the total bit-flip rate becomes dominant.
Instead, we observe that this transition point depends on the squeezing parameter $r$, 
with the transition point shifting towards larger values of $n_{\text{th}}$ for increasing values of $r$.
This becomes more apparent for larger values of $\alpha$, see \sfigref{fig:gain_suppression}{b} and \sfigref{fig:gain_suppression}{c}.
This behavior can be understood by associating the SCQ with an ordinary cat with exponentially larger amplitude $\beta = \alpha e^{r}$, leading to exponentially deeper potential wells that postpone the transition into the regime dominated by thermal noise.
In that respect, our results show a decreased sensitivity of the SCQ against thermal noise in comparison to the ordinary cat qubit in an experimentally relevant parameter regime of $n_{\text{th}} \approx \num{e-2}$. For instance, a moderately squeezed SCQ ($r=0.3$) shows no performance loss in contrast with the ordinary cat qubit which has a notably increased bit-flip rate $\GBF$, cf., \sfigref{fig:gain_suppression}{c}.
An interesting observation is that the exponential suppression factor $\gamma$ is not affected by thermal noise if $r \geq 0.3$ as apparent from \sfigref{fig:gain_suppression}{d}.

}

\subsubsection{Susceptibility to coherent Kerr evolution}
Any realistic implementation of the dissipative stabilization scheme in a superconducting circuit architecture will possess some form of undesired (Kerr) nonlinearity.
For the ordinary cat qubit, this does not pose an inherent problem since a combined stabilization scheme exists~\cite{gautier_combined_2022}.
Nevertheless,
even though it is possible to implement a combined stabilization scheme consisting of the dissipator $\mathcal{D}[\hat{b}^2 - \beta^2]$ and the squeezed Kerr Parametric Oscillator $\hat{H}_{\rm sKPO} \propto (\hat{b}^{\dagger 2} - \beta^2)(\hat{b}^2 - \beta^2)$, doing so would require of a higher order nonlinearity and
the activation of multiple non-resonant photon exchange processes as we detail in \appref{app:kerr_squeezed_cat}.

Here, we are interested in the effective bit-flip rate for the stabilized SCQ in the presence of a coherent evolution under the Kerr Hamiltonian $\hat{H}_{K} = K \hat{a}^{\dagger 2} \hat{a}^2$ in the general dynamics described by Eq.~\eqref{eq:master_eq_numerical_results}.
To this end, we consider single-photon losses with rate $\kappa_{-} / \kappa_{2} = 10^{-3}$ while neglecting pure dephasing $(\kappa_{\phi}=0)$ and single-photon gain $(\kappa_{+}=0)$.

\begin{figure}[!t]
    \centering
    \includegraphics{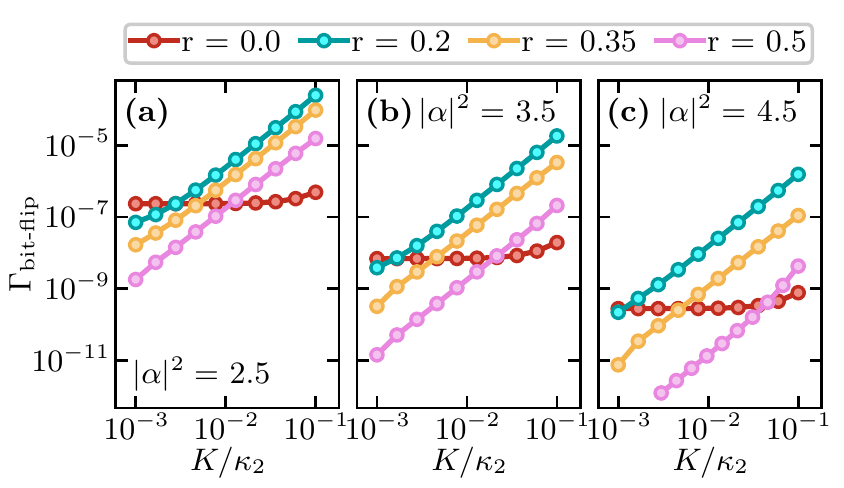}
\caption{Bit-flip error rate $\GBF$ in the presence of an undesired, coherent Kerr evolution. Each panel shows $\GBF$ for different values of the squeezing parameter $r$ and for a fixed value of the displacement $\alpha$. Results are obtained from simulating Eq.~\eqref{eq:master_eq_numerical_results} with $\kappa_{-} / \kappa_{2} = 10^{-3}$ and $\kappa_{\phi} = \kappa_{+} = 0$.}
    \label{fig:kerr_bit_flip_plot}
\end{figure}

The results of this section are summarized in \figref{fig:kerr_bit_flip_plot} where we plot the effective rate $\GBF$ as a function of the Kerr coupling $K$ for various values of the displacement amplitude $\alpha$ and squeezing parameter $r$.
The main observation from \figref{fig:kerr_bit_flip_plot} is that 
for fixed squeezing $r$ and displacement $\alpha$, we can identify a threshold value of the Kerr nonlinearity above which the cat qubit outperforms the SCQ.
More precisely, the threshold is defined as the Kerr coupling where the bit-flip rate of the cat code with code parameter $\alpha$, exceeds the bit-flip rate of the squeezed cat code with code parameters $(\alpha,r)$.
As the definition implies, this threshold depends on both the (achievable) displacement amplitude $\alpha$ as well as the (stabilized) squeezing parameter $r$.
A relevant observation is that for $r=0.2$ this threshold seems to become smaller with increasing $\lvert \alpha \rvert^2$, i.e., less Kerr can be tolerated, while for $r=0.35$ it stays approximately constant, and for $r=0.5$ the threshold increases with increasing $\lvert \alpha \rvert^2$.

These results indicate strong requirements for the effective Kerr nonlinearity of the device similar to the case for the dissipative stabilization of GKP states~\cite{campagne-ibarcq_quantum_2020}, likely with $\lvert K / \kappa_{2} \rvert \leq \num{e-2}$, which is not surprising given that in the infinite squeezing limit $(r \rightarrow \infty)$ the SQC attains similar properties as the GKP code.

\subsubsection{Bias-preserving $Z$ gate}
In this section we investigate the performance of the dissipative $Z$ gate for the SCQ.
Notice that one obtains a gate $G_{\rm SCQ}$ of the SCQ with code parameter $(\alpha, r)$ by applying the squeezing operator to the corresponding gate $G_{\rm cat}$ of the ordinary cat code with code parameter $\alpha e^{r}$, that is, $G_{\rm SCQ} = \hat{S} G_{\rm cat} \hat{S}^{\dagger}$.
A rotation around the $Z$ axis can be implemented via a resonant microwave drive with real amplitude $\epsilon_Z$, that is, the generator of a displacement along the $\hat{p}$ quadrature.
This would lead to an actual displacement in the absence of the dissipative stabilization scheme $(\kappa_{2} = 0)$, however, if $\epsilon_Z / \kappa_2$ is small, the combined evolution described by the following master equation
\begin{align}
    \label{eq:Z_gate_simulation_eq}
    \begin{split}
    \dv{\hat{\rho}}{t} =& -i \comm*{\epsilon_{Z} \hat{a}^{\dagger} + \epsilon_Z^{*} \hat{a}}{\hat{\rho}}  + \kappa_2 \mathcal{D}(\hat{b}^2 - \beta^2) \hat{\rho},
    \end{split}
\end{align}
results in a controlled rotation around the $Z$ axis of the Bloch sphere.
Here, the Hamiltonian part $\hat{H}_{Z} = \epsilon_{Z} \hat{a}^{\dagger} + \epsilon_Z^{*} \hat{a}$ is due to the resonant microwave drive applied to the oscillator.
Notice that the Hamiltonian does not change for SCQ apart from an exponential factor, that is, $\hat{S}(r) \hat{H}_{Z} \hat{S}^{\dagger}(r) = e^{r} \hat{H}_{Z}$, 
Hence, to implement the $Z$ for SCQ one applies the same resonant drive.
Initially, one might assume that the gate requires an exponentially stronger drive, however, given a desired rotation angle $\theta$ after a time $T_{\text{gate}}$, one chooses the following drive amplitude 
\begin{align}
    \label{eq:Z_gate_drive_amplitude}
    \epsilon_{Z} = \frac{\theta}{4 {\text Re}(\alpha) T_{\text{gate}}},
\end{align}
which is independent of the squeezing parameter $r$ because the exponential factor from the squeezing transformation above cancels with the exponential factor obtained from the replacement $\alpha \mapsto \beta = \alpha e^r$.

For a purely dissipative gate, one expects that the combined phase flip error is the result of an interplay of idling and non-adiabatic error $p_Z^{\text{NA}}$ ~\cite{gautier_combined_2022}
\begin{align}
    \label{eq:Z_gate_phase_error}
    p_Z = \kappa_{-} \lvert \alpha \rvert^2 T_{\text{gate}} + p_Z^{\text{NA}},
\end{align}
with $p_Z^{\text{NA}}$ given by
\begin{align}
    \label{eq:Z_gate_non_adiabatic_phase_error}
    p_Z^{\text{NA}} = \frac{\pi^2}{16 \lvert \alpha \rvert^4 \kappa_2 T_{\text{gate}}} e^{- 4 r},
\end{align}
which decreases with $e^{- 4 r}$.
For any finite value of $\kappa_{-}$ this interplay results in an optimal gate time $T_{\text{gate}}^{\text{opt}}$ which minimizes the combined phase flip error $p_Z$. 
For the SCQ this optimal gate time is given by
\begin{align}
    \label{eq:Z_gate_opt_gate_time}
    T_{\text{gate}}^{\text{opt}} = \frac{\pi}{4 \lvert \alpha \rvert^3 \sqrt{\kappa_{-} \kappa_{2} }} e^{- 2 r}.
\end{align}
Thus the $Z$ gate is exponentially faster in the squeezing parameter $r$.

\begin{figure}[t]
    \centering
    \includegraphics{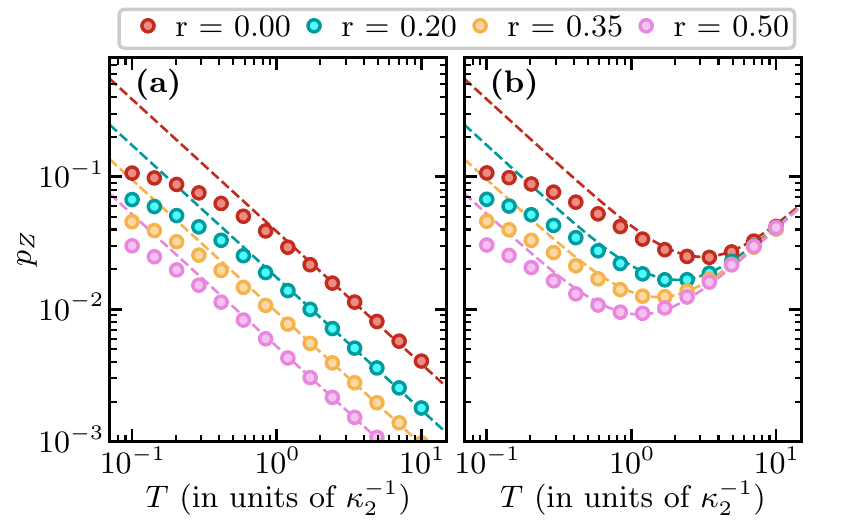}
    \includegraphics{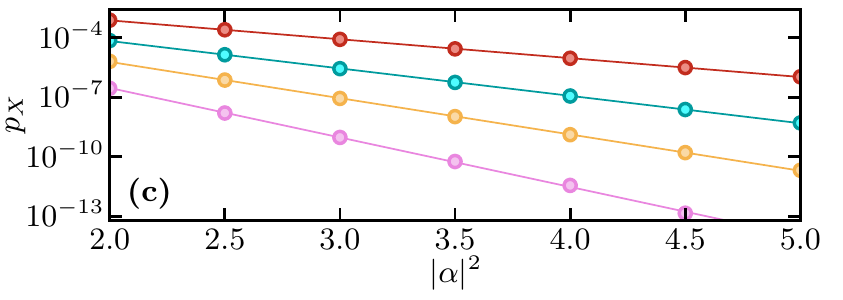}
    \caption{Gate induced phase and bit error for the $Z$ gate for $\lvert \alpha \rvert^2 = 4$ and various values of the squeezing parameter $r$.
    \textbf{(a)-(b)} Phase error as a function of the gate time $T$ 
    where the two panels correspond to different  $\kappa_{-} / \kappa_{2}$ ratios, in 
    \textbf{(a)} $\kappa_{-} / \kappa_{2} = 0$, i.e., phase errors occur due to non-adiabaticity of the gate, $p_Z = p_Z^{\text{NA}}$.
    \textbf{(b)} $\kappa_{-} / \kappa_{2} = \num{e-3}$. Markers represent numerical data obtained from the time evolution~\eqref{eq:Z_gate_simulation_eq} while dashed lines show theory results obtained from Eq.~\eqref{eq:Z_gate_phase_error}.
    \textbf{(c)} The bit-flip error probability $p_X$ after the $Z$ gate with optimal gate time $T_{\text{gate}}^{\text{opt}}$ and single-photon loss rate $\kappa_{-} / \kappa_{2} = \num{e-3}$. Markers show numerical data and solid lines are exponential fits.
    }
    \label{fig:Z_gate_phase_error}
\end{figure}

We have verified the above expressions by numerically simulating Eq.~\eqref{eq:Z_gate_simulation_eq} with an additional dissipator $\kappa_{-} \mathcal{D}[\hat{a}]$ accounting for single photon losses.
Our results are summarized in \sfigref{fig:Z_gate_phase_error}{a-b} where we have chosen a displacement amplitude $\alpha$ such that $\lvert \alpha \rvert^2 = 4$ and have compared the two cases $\kappa_{-} / \kappa_{2} = 0$ and $\kappa_{-} / \kappa_2 = \num{e-3}$ to separate out the effect of idling errors in the first case. 
We see that the analytical model~\eqref{eq:Z_gate_phase_error} (dashed lines) reproduces the numerical data (markers) accurately.
We emphasize that even though the here selected range for the squeezing parameter $r$ is moderate, see also \sfigref{fig:squeezed_cat_code}{d}, the SCQ allows a reduction in the $p_Z$ error by roughly an order of magnitude in comparison with the cat qubit while also reducing the optimal gate time to a third of it.
Importantly, also for the SCQ our implementation of the $Z$ gate remains bias-preserving.
This can be seen in \sfigref{fig:Z_gate_phase_error}{c} where we show the bit-flip error probability $p_X$ after the $Z$ gate with optimal gate time $T_{\text{gate}}^{\text{opt}}$ and single-photon loss rate $\kappa_{-} / \kappa_{2} = \num{e-3}$.
The solid lines show exponential fits to the numerical data and we find that the slope scales approximately as $\exp(2r)$.

The main takeaway message from this section is that the increased performance of the SCQ over the ordinary cat qubit remains during the $Z$ gate operation.
In particular, we find that the existing model for the phase error $p_Z$ of the cat qubit can be easily adapted to the SCQ by a replacement $\alpha \mapsto \alpha e^{r}$ with some caution, for example, in the first term in Eq.~\eqref{eq:Z_gate_phase_error} this substitution is not done because the phase-flip rate due to single-photon losses is independent of the squeezing parameter $r$, see \sfigref{fig:squeezed_cat_code}{c}. 
We omit here the analysis of multi-qubit gates, such as the \qgate{CNOT}, since the numerical simulations become too demanding.
However, we do expect similar results as for the $Z$ gate, that is, an exponential suppression of errors in the squeezing parameter $r$.

In the future, it would be interesting to combine the dissipative gate implementation with a squeezed version of the two-photon exchange Hamiltonian originally proposed in Ref.~\cite{gautier_combined_2022} as well as including derivative-based corrections in the envelope of $\epsilon_Z$ as proposed in Ref.~\cite{xu_engineering_2022} for the cat code.


\subsection{Circuit QED Implementation \label{ssec:circuit_qed_implementation}}
\begin{figure}[t]
    \centering
    \includegraphics{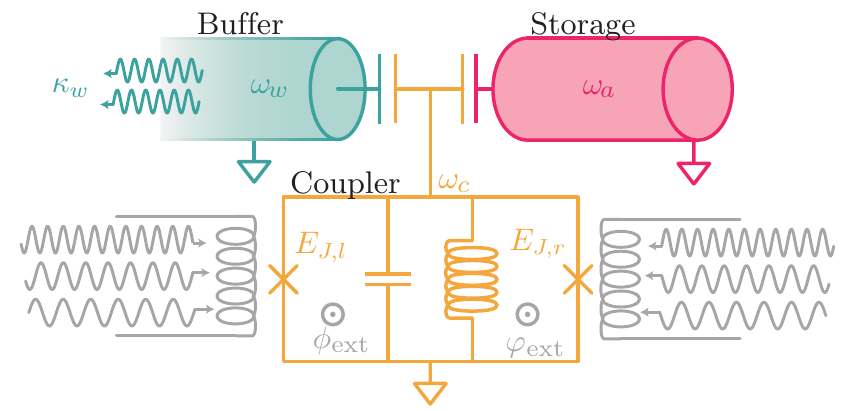}
    \caption{Sketch of a possible circuit QED implementation of the squeezed cat confinement scheme. 
    The storage resonator (red) that hosts the stabilized squeezed cat qubit and the buffer mode (teal) are coupled capacitively through a non-linear coupler (orange) made from an asymmetric Josephson-Junction loop to enable flux-pumped three-wave mixing.
    By biasing the coupler loops with the external fluxes $\phi_{\rm ext}$ and $\varphi_{\rm ext}$ satisfying the relations in Eq.~\eqref{eq:flux_bias}, residual Kerr interactions can be dramatically minimized.
    }
    \label{fig:circuit_sketch}
\end{figure}

Here we propose a possible superconducting circuit implementation for the dissipative stabilization of the squeezed cat code.
Our scheme is based on a modification of the protocol by Lescanne \emph{et al.}~\cite{lescanne_exponential_2020} for the dissipative stabilization of the ordinary cat qubit. 
It utilizes a high-quality resonator referred to as the storage, coupled via a three-wave mixing element to a lossy auxiliary resonator referred to as the buffer.
Recall that to engineer the effective dissipator $\mathcal{D}(\hat{b}^2 - \beta^2)$ for the storage resonator, we first need to generate a resonant interaction between the storage and buffer resonators of the form
\begin{align}
    \label{eq:three_wave_mixing_interaction}
    \hat{H}_{\mathrm{int}} &= g_3 \left( \hat{w}^{\dagger} \hat{b}^2 + \hat{w} \hat{b}^{\dagger 2} \right) \\
    &= g_3 \hat{w}^{\dagger} \left(\cosh(r) \hat{a} - \sinh(r) \hat{a}^{\dagger}\right)^2 + \mathrm{H.c.},
\end{align}
where $\hat{w}$ and $\hat{a}$ are the annihilation operators of the buffer and storage resonator, respectively. 
We propose to achieve such an interaction with the superconducting circuit shown in \figref{fig:circuit_sketch}.
For the coupler, we consider an asymmetrically threaded SQUID (ATS) for its three-wave mixing capabilities.
The Hamiltonian of the sketched superconducting circuit in \figref{fig:circuit_sketch} takes the form~\cite{hillmann_designing_2022}
\begin{align}
    \label{eq:ATS-Hamiltonian-general}
    \hat{H} &= \omega_{a} \hat{a}^{\dagger} \hat{a} + \omega_c \hat{c}^{\dagger} \hat{c} + \omega_{w} \hat{w}^{\dagger} \hat{w} \nonumber \\
    &- 2 E_J \left[\cos(\varphi_{\Sigma}) \cos(\hat{\varphi}  + \varphi_{\Delta}) \right],
\end{align}
where $\hat{c}$ is the annihilation operator corresponding to the coupler mode.
Furthermore, $\omega_a$, $\omega_c$, and $\omega_w$ correspond to the frequencies of the respective modes, while $\hat{\varphi}$ describes the hybridized mode in the coupler and is given by $\hat{\varphi} = (\varphi_c \hat{c} + \varphi_a \hat{a} + \varphi_{w} \hat{w} + \mathrm{H.c.})$ where $\varphi_{x}$ 
is
the participation ratio of the respective mode in the Josephson junction which will depend on the macroscopic parameters of the circuit elements.
We have also introduced $\varphi_{\Sigma} = \varphi_{\mathrm{ext}} + \phi_{\mathrm{ext}}$ and $\varphi_{\Delta} = \varphi_{\mathrm{ext}} - \phi_{\mathrm{ext}}$ which correspond to the sum and difference of the external fluxes applied through the loops, respectively.
To engineer the desired interaction we operate the coupler with flux biases
\begin{align}
    \label{eq:flux_bias}
    \varphi_{\Sigma} = \frac{\pi}{2} + \varphi_{\Sigma}^{\mathrm{ac}}(t), \quad \varphi_{\Delta} = \frac{\pi}{2},
\end{align}
where $\varphi_{\Sigma}^{\mathrm{ac}}(t)$ is an additional three tone flux modulation 
of the form 
\begin{align}
    \label{eq:ac_flux_drive}
    \varphi_{\Sigma}^{\mathrm{ac}}(t) &= \epsilon_1 \cos(\omega_1 t) + \epsilon_2 \cos(\omega_2 t) + \epsilon_3 \sin(\omega_3 t),
\end{align}
with $\lvert \varphi_{\Sigma}^{\mathrm{ac}}(t) \rvert \ll 1$.
At the chosen flux bias the inductive energy of the ATS becomes anti-symmetric and the Hamiltonian~\eqref{eq:ATS-Hamiltonian-general} reduces to
\begin{align}
    \label{eq:Hamiltonian_at_flux_bias}
    \hat{H} = \omega_{a} \hat{a}^{\dagger} \hat{a} + \omega_c \hat{c}^{\dagger} \hat{c} + \omega_{w} \hat{w}^{\dagger} \hat{w} - 2 E_J \varphi_{\Sigma}^{\mathrm{ac}}(t) \sin(\hat{\varphi}).
\end{align}
To obtain the desired $\hat{\varphi}^3$ interaction we expand the sine up to third order in $\hat{\varphi}$ leading to 
\begin{align}
    \hat{H} &= \omega_{a} \hat{a}^{\dagger} \hat{a} + \omega_c \hat{c}^{\dagger} \hat{c} + \omega_{w} \hat{w}^{\dagger} \hat{w} \nonumber \\
    &- 2 E_J \left[ \varphi_{\Sigma}^{\mathrm{ac}}(t) \hat{\varphi} - \varphi_{\Sigma}^{\mathrm{ac}}(t) \frac{\hat{\varphi}^3}{6} \right].
\end{align}
We can further eliminate the term linear in $\hat{\varphi}$ by moving to a 
time-dependent joint
displaced frame specified by the displacements
\begin{align}
    \xi_x(t) =\sum_{k=1}^{3} \frac{-i E_J \varphi_x \epsilon_k}{i\left(\omega_x-\omega_{k}\right)+\kappa_x / 2} e^{-i \omega_{k} t}, \quad x=a, c, w,
\end{align}
where $\kappa_a$, $\kappa_c$, and $\kappa_w$ denote the single-photon loss rates of the storage, coupler, and buffer mode, respectively.
As a result $\hat{\tilde{\varphi}}$ now takes the form
\begin{align}
    \hat{\tilde{\varphi}} = [ \varphi_c( \hat{c} + \xi_{c}) + \varphi_c( \hat{a} + \xi_{a}) + \varphi_c( \hat{w} + \xi_{w}) + \mathrm{H.c.}],
\end{align}
leading to the Hamiltonian
\begin{align}
    \hat{H} &= \omega_{a} \hat{a}^{\dagger} \hat{a} + \omega_c \hat{c}^{\dagger} \hat{c} + \omega_{w} \hat{w}^{\dagger} \hat{w} 
    + E_J \varphi_{\Sigma}^{\mathrm{ac}}(t) \frac{\hat{\tilde{\varphi}}^3}{3}.
\end{align}
To obtain from this Hamiltonian the desired interaction Hamiltonian~\eqref{eq:three_wave_mixing_interaction}, we choose the pump frequencies $\omega_i$ such that $\omega_1 = 2\omega_a - \omega_w$, $\omega_2 = 2 \omega_a + \omega_w$, $\omega_3 = \omega_w$.
This choice is necessary to enable all interactions in Eq.~\eqref{eq:three_wave_mixing_interaction}.
Then, in order to obtain the right coefficients the flux-pump amplitudes are chosen such that $\epsilon_1(r) = \lambda \cosh^2(r)$, $\epsilon_2(r) = \lambda \sinh^2(r)$, and $\epsilon_3(r) = \lambda \sinh(2r)$ with $\lambda \ll 1$ the strength of the drive tone.
%
%
%

Furthermore, for us to assume that the coupler stays in its ground state, i.e., $\expval{\hat{c}^{\dagger} \hat{c}} = 0$, we require that the frequency of the coupler mode $\omega_c$ is sufficiently detuned from all of the pump frequencies.

Note that the pump tone at $\omega_3 = \omega_w$ will additionally lead to a Hamiltonian term acting on the waste mode that is given by
\begin{align}
    \label{eq:drive_hamiltonian}
    \hat{H}_{\mathrm{dr}} = \tilde{\Omega}^* \hat{w} + \tilde{\Omega} \hat{w}^{\dagger},
\end{align}
where $\tilde{\Omega}$ describes the strength of this effective linear drive on the waste mode given by
\begin{align}
    \tilde{\Omega} = \sum_{x=a, c, w} \frac{i E_J \epsilon_3(r) \varphi_x^2}{i\left(\omega_x-\omega_{3}\right)+\kappa_x / 2} \approx \frac{i E_J \epsilon_3(r) \varphi_{w}^2}{\kappa_{w}/2}.
\end{align}


Under the nonlinear interaction term in Eq.~\eqref{eq:three_wave_mixing_interaction}, the photons injected in the waste mode by this drive are converted into superpositions of displaced squeezed states.
Thus, this term is necessary to stabilize a squeezed cat state with non-zero amplitude $\alpha \neq 0$.
However, since the amplitude $\tilde{\Omega}$ of this effective drive is not an independent parameter, that is, it is determined by the macroscopic properties of the superconducting circuit and the flux pump amplitude $\epsilon_3(r)$, we consider adding an additional charge drive of the form~\eqref{eq:drive_hamiltonian}.
\timo{Both the driving Hamiltonian activated through the flux pump and the additional one applied through the charge line are described by the same Hamiltonian, but with different amplitudes. By absorbing the amplitude of the charge pump into $\tilde{\Omega}$, we can define a renormalized driving amplitude, denoted by $\Omega$. This allows us to treat both sources of driving with a single effective Hamiltonian.
Then, to prepare a squeezed cat state $\ket{\mathcal{C}_{\alpha, r}^{+}}$ with amplitude $\alpha$ and squeezing $r$, one chooses the amplitude of the the charge line drive such that the effective driving amplitudes becomes $\Omega = - g_3 \alpha^2 \exp(2  r)$. 
}

To obtain an effective single-mode description of the form given by Eq.~\eqref{eq:ideal_cat_code_dissipator} from the previously described system and drive Hamiltonian, we assume that the waste mode $w$ exhibits strong single-photon dissipation with rate $\kappa_{w}$ described by the dissipator $\kappa_{w} \mathcal{D}[\hat w]$.
In the limit where $\kappa_{w} \gg g_3 \cosh r$ one can adiabatically eliminate the waste mode to obtain the effective two-photon dissipator $\kappa_{2} \mathcal{D}[\hat{b}^2 - \beta^2]$ with rate $\kappa_{2} = 4 g_3^2 / \kappa_{w}$.
As described in Ref.~\cite{chamberland_building_2022} the validity of this approximation will also depend on the size of the (squeezed) cat state that one aims to prepare.
For finite $\kappa_w$, the additional condition $2 \lvert \alpha \rvert g_3 \ll \kappa_{w}$ arises.
In practice numerical master equation simulations have shown that it is sufficient to achieve $2 \lvert \alpha \rvert g_3 / \kappa_{w} < 1/5$.
This condition for the adiabatic elimination remains in the case of squeezed cat states.

Lastly, let us mention that a finite asymmetry $\eta = (E_{J, l} - E_{J, r}) / (E_{J, l} + E_{J, r})$ of the junction energies, e.g., due to fabrication variances,  will lead to an additional term $- 2 E_J \eta \cos(\hat{\varphi})$ in the Hamiltonian~\eqref{eq:Hamiltonian_at_flux_bias} which will lead to static Kerr-type nonlinearities in the Hamiltonian of the storage mode.
As shown in \figref{fig:kerr_bit_flip_plot} these nonlinearities significantly increase the effective bit-flip rate of the squeezed cat qubit.
For example, in the experiment by Lescanne \emph{et al.}~\cite{lescanne_exponential_2020}, the ratio of the measured Kerr coupling $K$ and two-photon dissipation rate $\kappa_{2}$ was $\lvert K / \kappa_{2} \rvert \approx 1/5$.
We expect that the resulting nonlinearity could be reduced by operating the device at an altered dc-flux-bias to take advantage of Kerr renormalization through the $\varphi^3$ interaction~\cite{petrescu_accurate_2021-1}.

\section{Discussion and Conclusion\label{sec:discussion}}
Bosonic codes offer a hardware-efficient way of redundantly encoding logical information into a subspace of a much larger Hilbert space.
Furthermore, with the recent surge in interest of quantum error-correcting that are tailored to a noise-bias in the effective Pauli error model, biased-noise bosonic qubits such as the cat qubit have received considerable experimental~\cite{lescanne_exponential_2020, berdou_one_2022, grimm_stabilization_2020} and theoretical~\cite{puri_stabilized_2019, gautier_combined_2022, guillaud_repetition_2019} attention.
However, while the cat qubit exhibits a large error bias, the exponential reduction of bit-flip errors comes with the caveat of a linearly increasing phase-flip rate.
To overcome this shortcoming, recently Schlegel, Minganti and Savona introduced the squeezed cat qubit (SCQ)~\cite{schlegel_quantum_2022} that in principle allows further suppression of the bit-flip rate at a fixed phase-flip rate by increasing the peak squeezing.

In this article, we have proposed a dissipative stabilization mechanism for squeezed cat states and analyzed the error correction performance of the squeezed cat qubit, the quantum error correcting code derived from squeezed cat states, in a superconducting circuit-inspired error model.
To this end, we have performed numerical simulations for various realistic noise models to extract the effective phase- and bit-flip error rates within that model and compared the results to the ordinary, non-squeezed cat qubit.
We have found that for all analyzed incoherent noise processes, the bit-flip error rates of the SCQ are exponentially lower in the squeezing parameter $r$ in comparison to the non-squeezed cat qubit, while phase-flip error rates are approximately independent of the squeezing parameter.
Thus, the squeezing allows further increasing the noise bias of the encoded qubit which is relevant for hardware-efficient fault-tolerant quantum computation using bias-tailored quantum error-correcting codes which have seen a lot of interest recently.
Importantly, our results further suggest that the increased noise bias is preserved during gate operation, that is, we find that the $Z$ gate on the SCQ can be performed exponentially faster and with exponentially lower phase error probability than on the cat qubit, while at the same time significantly reducing the bit-flip error probability.
However, we identify a susceptibility to undesired coherent Kerr evolution, imposing strong requirements on the residual nonlinearity of the bosonic mode which hosts the stabilized squeezed cat qubit.
While we believe that residual nonlinearities can be significantly reduced by carefully taking into account terms beyond the rotating wave approximation~\cite{hillmann_designing_2022}, it is an open question whether alternative approaches exist.
We conclude by noting that our suggested implementation of the engineered dissipative dynamics within a superconducting circuit platform is readily realizable in state-of-the-art devices~\cite{lescanne_exponential_2020} and allows for tuning the squeezing parameter in situ. In fact, this is exactly the same setup currently used for the stabilization of cat states in which additional modulation drives allow the parametrical activation of the full cavity squeezed mode operator. We believe that, while theoretically possible, the Hamiltonian confinement of a SCQ is challenging from an experimental point of view. The study of its plausible implementation in superconducting circuits is left for future work. 

\textit{Note added}\textemdash
While writing this manuscript, we became aware of a similar work analyzing the performance of dissipatively stabilized squeezed cat qubits~\cite{xu_autonomous_2022}.

\acknowledgments
T.H. acknowledges the financial support from the Chalmers Excellence Initiative Nano and the Knut and Alice Wallenberg Foundation through the Wallenberg Centre for Quantum Technology (WACQT).
F.Q. acknowledges support from the Okinawa Institute of Science and Technology Graduate University.

\appendix
\section{Unconditional state preparation}
The protocol described in this work requires the initialization of the system (storage) into a parity eigenstate, i.e., Fock states or SCQs. While this can be routinely done in circuit QED it might not be the case for different quantum technologies. 
Alternatively, here we will show how to stabilize an even parity squeezed cat state 
unconditionally, that is, without the need for initialization.

Consider a general non-Hermitian operator $\hat{L}$ with a unique dark state or zero eigenvalue eigenstate $\ket{\psi}$, i.e., $\hat{L} \ket{\psi} = 0$.
The steady-state $\partial_t \hat{\rho}_{\rm ss} = 0$ of the dissipative dynamics $\partial_t \hat{\rho} = \kappa \mathcal{D}[\hat{L}] \hat{\rho}$ will therefore correspond to the dark state of $\hat{L}$: $\hat{\rho}_{\rm ss} = \ket{\psi} \bra{\psi}$. The latter is true regardless of the initial state of the system. 
Two well-known examples correspond to the case of a reservoir at zero-temperature which cools a resonator to the photon vacuum state $\ket{0}$ where $\hat{L} = \hat{a}$, and the cooling of a resonator to a squeezed vacuum state $\ket{\xi} = \hat{S}(\xi) \ket{0}$, where $\hat{L} = \mu \hat{a} + \nu \hat{a}^\dagger$, with $\abs{\mu}^2 - \abs{\nu}^2 = 1$~\cite{dassonneville_dissipative_2021}.

As shown in Ref.~\cite{mamaev_dissipative_2018}
the dark state of the nonlinear operator
\begin{equation}
    \hat{L} = \left( \mu_0 + \mu_1 \hat a^\dagger \hat a \right) \hat a + \nu \hat a^\dagger,
\end{equation}
approaches asymptotically 
an even parity (two-headed) cat state
of amplitude $\alpha = i \sqrt{\nu/ \mu_1}$,
i.e., $\hat{L} \ket*{\mathcal{C}^+_{\alpha}} = 0$ in the limit $\mu_0/ \mu_1 \to 0$. By means of a unitary transformation $\hat U$, we can find the annihilator of the state $\hat{U} \ket*{\mathcal{C}^+_{\alpha}}$
\begin{equation}
    \left( \hat U \, \hat z \,\hat U^\dagger \right) \left( \hat U \ket*{\mathcal{C}^+_{\alpha}} \right) = 0 .
\end{equation}
By choosing $\hat U \equiv \hat{S}(\xi)$ we have $\hat U \ket{C^+_\alpha} = \mathcal{N}( \ket{\alpha}_{\xi} + \ket{-\alpha}_{\xi})$, with $\mathcal{N}$ a normalization factor.  
Using the relation $\hat{S}(\xi) \hat{D}(\alpha) = \hat{D}(\beta_{\alpha, \xi}) \hat{S}(\xi)$, with
$\beta_{\alpha, \xi} = \alpha \cosh(r) - \alpha^* {\rm e}^{-i \phi} \sinh(r)$, we have that 
$\hat U \ket{C^+_\alpha} = \ket*{\mathcal{C}^+_{\beta_{\alpha, \xi}, \xi}}$
and thus we arrive at our desired relation
\begin{align}
    \hat X \ket*{\mathcal{C}^+_{\beta_{\alpha, \xi}, \xi}} = 0,
\end{align}
with
\begin{align}
    \hat X  = \hat S(\xi) \left[ \left( \mu_0 + \mu_1 \hat a^\dagger \hat a \right) \hat a + \nu \hat a^\dagger \right]  S^\dagger(\xi).
\end{align}
Expanding the above expression we get
\begin{widetext}
\begin{align}
    \hat X &=  \left[ \mu_0 \cosh(r) + \nu {\rm e}^{-i \phi} \sinh(r) + 3 \mu_1 \sinh^2(r) \cosh(r) \right] \hat a \nonumber\\
    &+ \left\{ \mu_0 {\rm e}^{i \phi} \sinh(r) + \nu \cosh(r) + {\rm e}^{i \phi} \sinh(r) \left[ \cosh(2r) + \sinh^2(r) \right] \right\} \hat a^\dagger  \nonumber\\
    &+  \mu_1 \cosh(r) \left[ \cosh(2r) + \sinh^2(r) \right] \hat a^\dagger \hat a^2 \nonumber\\
    &+ \mu_1 {\rm e}^{i \phi} \sinh(r) \left[ \cosh^2(r) + \cosh(2r) \right] \hat a^{\dagger 2} \hat a \nonumber\\
    &+ \mu_1 {\rm e}^{-i \phi} \sinh(r) \cosh^2(r) \hat a^3 + \mu_1 {\rm e}^{2i \phi} \sinh^2(r) \cosh(r) \hat a^{\dagger 3} .
\end{align}
\end{widetext}
Therefore, the steady-state of the Lindblad master equation $\partial_t \hat{\rho} = \kappa \mathcal D[\hat X] \hat{\rho}$ is $\hat{\rho}_{\rm ss} = \ket*{\mathcal{C}^+_{\beta_{\alpha, \xi}, \xi}} \bra*{\mathcal{C}^+_{\beta_{\alpha, \xi}, \xi}}$.
 
\section{Hamiltonian Confinement Scheme for the Squeezed Cat Qubit \label{app:kerr_squeezed_cat}}

A different approach to initialize and confine a cat qubit utilizes a parametrically driven Kerr oscillator or Kerr parametric oscillator (KPO) described by the Hamiltonian
\begin{align}\label{eq:kpo}
    \hat{H} = K \hat{a}^{\dagger 2} \hat{a}^2 + \epsilon_2 \left( \hat{a}^{\dagger 2} + \hat{a}^2 \right) .
\end{align}
By noticing that the latter can be rewritten as 
$\hat{H} =  K(\hat{a}^{\dagger 2} - \alpha^2) (\hat{a}^2 - \alpha^2) + K \alpha^4$, with $\alpha = \sqrt{\epsilon_2 / K}$, it is straightforward to see that the coherent states $\ket{\pm \alpha}$ or equivalently, the cat states $\ket*{\mathcal{C}^{\pm}_\alpha}$ are degenerate eigenstates of the KPO Hamiltonian. Furthermore, the cat manifold is protected from the rest of the spectrum by an energy gap~\cite{puri_engineering_2017, puri_stabilized_2019, grimm_stabilization_2020}. 

An advantage of this approach compared to the dissipative stabilization is that the KPO Hamiltonian is rather simple to realize in experiments. It consists of a linear resonator coupled to a dc SQUID or a SNAIL, flux or current modulated at twice the frequency of the resonator mode respectively. Furthermore, this stabilization scheme is compatible with fast gates which maintain the noise bias of the cat code~\cite{xu_engineering_2022}. In addition, it has been recently shown that both, dissipative and Hamiltonian confinement can be merged in a single platform with an improved gate performance as compared to each individual scheme~\cite{gautier_combined_2022}.

One could argue that all of the above ideas can be straightforwardly extended to the SCQ. Indeed this is the case, by squeezing Hamiltonian \eqref{eq:kpo} we arrive to the squeezed Kerr parametric oscillator (sKPO) Hamiltonian, $\hat{H}_{\rm sKPO} =\hat{S}(r) \hat{H} \hat{S}^\dagger (r) =  K \hat{b}^{\dagger 2} \hat{b}^2 + \epsilon_2 ( \hat{b}^{\dagger 2} + \hat{b}^2 )$, with $\hat{b} = \hat{S}(r) \hat{a} \hat{S}^\dagger (r)$ and squeezed cat states as its eigenstates (we have noticed some recent publications in which the authors refer to the KPO \eqref{eq:kpo} as the \emph{squeezed Kerr-nonlinear oscillator}. By \emph{squeezing} the authors refer to the two-photon drive.).

In principle, the full squeezed Kerr nonlinearity, i.e., $\hat{b}^{\dagger 2}\hat{b}^2$ could be activated by means of parametric drives. Expanding this nonlinear potential in terms of the storage bosonic operator reveals rotating terms of the form: $\hat{a}^{4}$, $\hat{a}^{\dagger 3}\hat{a}$, $\hat{a}^2$, together with their Hermitian conjugates and the non rotating Kerr nonlinearity $\hat{a}^{\dagger 2} \hat{a}^2$. In order to parametrically activate the rotating terms with frequencies $4\omega_a$ and $2 \omega_a$, a nonlinear interaction of at least order four is necessary using flux driving and at least of order five using current modulation. 
Regardless, while a modulation with frequency $2 \omega_a$ allows selecting quadratic and fourth-order terms from both nonlinear terms, it may be challenging to modulate the system at $4 \omega_a$. 
For typical values of resonator frequencies (few GHz), such a high-frequency modulation might excite the plasma frequency of the Josephson junctions, typically around \SI{20}{\GHz}.
In order, to avoid this, one alternative would be to design a resonator with a smaller frequency or utilize Josephson junctions based on constrictions instead of tunnel junctions.

\section{Numerical methods \label{app:numerical_methods}}
\subsection{Observables of the encoded state}
In order to accurately estimate the effective phase- and bit-flip rates of the squeezed cat code we relate them to the properties of the full quantum harmonic oscillator space.
Concretely we relate the expectation value of Pauli $X$ $\expval{\sigma_X} = \Trace[\hat{J}_{x} \hat{\rho}]$ where $\hat{J}_{x}$ is the Fock parity operator, that is, 
\begin{align}
    \hat{J}_x = \hat{J}_{++} - \hat{J}_{--},
\end{align}
with 
\begin{align}
    \hat{J}_{++} &= \sum_{n=0}^{\infty} \ketbra{2n},\\
    \hat{J}_{--} &= \sum_{n=0}^{\infty} \ketbra{2n + 1}.
\end{align}
On the other hand, the expectation value of Pauli $Z$  $\expval{\sigma_Z} = \Trace[\hat{J}_{Z} \hat{\rho}]$ whether the state has support in the positive or negative half-plane of the phase space.
Here we use an adaption of the observable introduced in Ref.~\cite{mirrahimi_dynamically_2014} which is a good approximation of $\mathop{\mathrm{sign}}(\hat{a} + \hat{a}^{\dagger})$ and is defined as 
\begin{align}
    \hat{J}_z = \hat{J}_{+-} + \hat{J}_{+-}^{\dagger},
\end{align}
with 
\begin{align}
    \label{eq:app_j_pm}
    \hat{J}_{+-}=\sqrt{\frac{2 \alpha^2}{\sinh \left(2 \alpha^2\right)}} \sum_{q=-\infty}^{\infty} \frac{(-1)^q}{2 q+1} I_q(\alpha^2) \hat{J}_{+-}^{(q)},
\end{align}
where $\alpha \in \mathbb{R}$, $I_q(x)$ is the modified Bessel function of the first kind and $\hat{J}^{(q)}_{+-}$ is further defined as
\begin{align}
\hat{J}_{+-}^{(q)}=\left\{\begin{array}{ll}
\frac{\left(\hat{a}^{\dagger} \hat{a}-1\right) ! !}{\left(\hat{a}^{\dagger} \hat{a}+2 q\right) ! !} \hat{J}_{++} \hat{a}^{2 q+1} \quad q \geq 0 \\
\hat{J}_{++} \hat{a}^{\dagger(2|q|-1)}   \frac{\left(\hat{a}^{\dagger} \hat{a}\right) ! !}{\left(\hat{a}^{\dagger} \hat{a}+2|q|-1\right) ! !} & q<0
\end{array}\right.,
\end{align}
where $n!! = (n-2)!! n$ denotes the double factorial.
While $\hat{J}_{z}$ also correctly captures the decay rate of the squeezed cat qubit, the SCQ basis states are eigenstates with eigenvalues less than one.
Thus, to correctly normalize $\hat{J}_z$, we instead compute $\hat{S}(r) \hat{J}_z \hat{S}^{\dagger}(r)$ while additionally replacing $\alpha^2$ with $\beta^2 = \alpha^2 e^{2 r}$ in $\hat{J}_{+-}$ [Eq.~\eqref{eq:app_j_pm}].

\subsection{Estimating error rates}

\begin{figure}[H]
    \centering
    \includegraphics{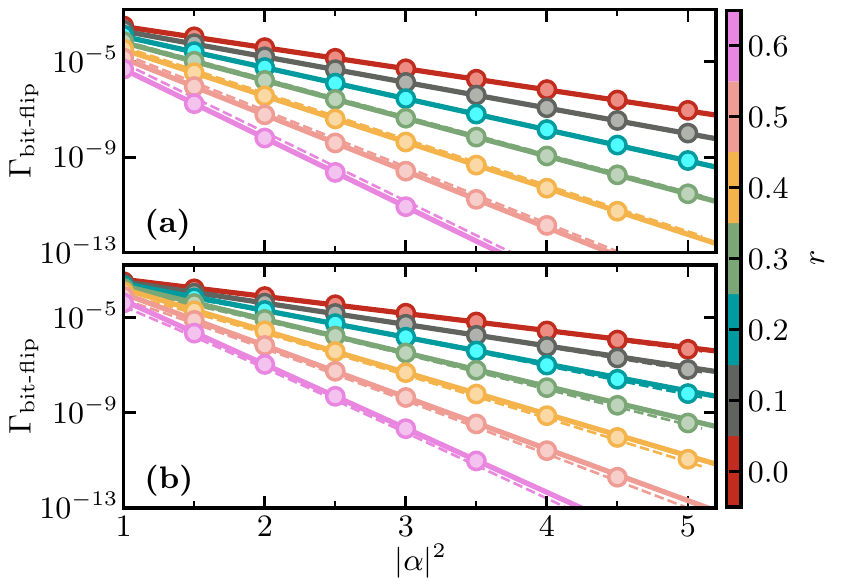}
    \caption{Exemplary data for pure dephasing and single-photon gain noise from which the exponential suppression factor $\gamma$ can be extracted. Markers show numerical data, solid lines are exponential fits to the data, and dashed lines correspond to our approximate theory Eqs.~\eqref{eq:bit_flip_rate_dephasing_thy} and~\eqref{eq:bit_flip_rate_gain_thy} for panels (a) and (b), respectively. \textbf{(a)} Bit-flip rate with pure dephasing and single-photon losses with rates $\kappa_{\phi} / \kappa_2 = \kappa_{-} / \kappa_2 = \num{e-3}$.
     \textbf{(a)} Bit-flip rate with single-photon loss and gain with rates $\kappa_{+} / \kappa_2 = \kappa_{-} / \kappa_2 = \num{e-3}$.}
    \label{fig:app_gain_dephasing_flip}
\end{figure}

We numerically calculate the effective bit- and phase-flip rates by fitting the decaying logical observables,
\begin{align}
   \Trace[\hat{J}_z \hat{\rho}(t)] &\sim \mathrm{e}^{-\GBF t}, \\
    \Trace[\hat{J}_x \hat{\rho}(t)] &\sim \mathrm{e}^{-\GPF t}.
\end{align}
To this end, we initialize the system in an ideal state $\ket*{\SCQ{0}{\beta}}$ or $\ket*{\SCQ{+}{\beta}}$ before time evolving it for a time $t \gg t_{\mathrm{conf}} \approx (4 \alpha^2 \kappa_2)^{-1}$ under the dynamics generated by the Lindbladian Eq.~\eqref{eq:master_eq_numerical_results}.
The simulations are performed using the \texttt{qutip} package~\cite{johansson_qutip_2013}, then the data is analyzed and visualized utilizing \textsc{python} libraries~\cite{hunter_matplotlib_2007, harris_array_2020, virtanen_scipy_2020}.
The exponential suppression factor $\gamma$ is obtained from fitting an exponential $\exp(- \gamma \alpha^2)$ to the bit-flip rate over a range of displacement amplitudes $2 \leq \alpha^2 \leq 5$ for all $\GBF \gtrapprox \num{e-13}$ which is our numerical accuracy threshold.
\timo{For bit-flip rates below this value observe numerical instabilities in our fitting routines.}
We show additional data for pure dephasing and single-photon gain noise in \figref{fig:app_gain_dephasing_flip}.

\end{document}